\documentclass[aps,prl,superscriptaddress]{revtex4}
\usepackage{amsmath,epsfig}
\usepackage{graphicx}

\def\be{\begin{equation}}
\def\ee{\end{equation}}
\def\bea{\begin{eqnarray}}
\def\eea{\end{eqnarray}}
\newcommand{\ket}[1]{\mbox{$|#1\rangle$}}
\newcommand{\bra}[1]{\mbox{$\langle#1|$}}
\newcommand{\avg}[1]{\mbox{$\langle#1\rangle$}}

\def\GammaD{\Gamma_{\footnotesize\textrm{1D}}}

\begin{document}

\title{Self-organization of atoms along a nanophotonic waveguide}
\date{\today}

\author{D.E. Chang}
\affiliation{ICFO - Institut de Ciencies Fotoniques, Mediterranean
Technology Park, 08860 Castelldefels (Barcelona), Spain}
\email{darrick.chang@icfo.es}

\author{J.I. Cirac}
\affiliation{Max-Planck-Institut f\"{u}r Quantenoptik, Hans-Kopfermann-Str. 1, D-85748 Garching, Germany}

\author{H.J. Kimble}
\affiliation{IQIM, California Institute of Technology, Pasadena,
CA 91125, USA} \affiliation{Norman Bridge Laboratory of Physics
12-33, California Institute of Technology, Pasadena, CA 91125,
USA}

\begin{abstract}
Atoms coupled to nanophotonic interfaces represent an exciting frontier for the investigation of quantum light-matter interactions. While most work has considered the interaction between statically positioned atoms and light, here we demonstrate that a wealth of phenomena can arise from the self-consistent interaction between atomic internal states, optical scattering, and atomic forces. We consider in detail the case of atoms coupled to a one-dimensional nanophotonic waveguide, and show that this interplay gives rise to self-organization of atomic positions along the waveguide, which can be probed experimentally through distinct characteristics of the reflection and transmission spectra.
\end{abstract}
\maketitle


In recent years, ultracold atomic gases have become rich systems for the investigation of novel many-body phenomena involving spatial and/or spin degrees of freedom~\cite{bloch_many-body_2008}. The success of this field takes advantage of the ability to trap atoms in optical lattices and manipulate their interactions, enabling one to controllably emulate more complex condensed matter systems. In the majority of these experiments, however, the underlying lattice is static and provided by external lasers, which prohibits one from accessing phenomena associated with the back-action of lattice dynamics on spatial and internal dynamics.

This potentially rich behavior has been explored recently using cold atoms coupled to Fabry-Perot optical cavities~\cite{domokos_collective_2002,black_observation_2003,gopalakrishnan_emergent_2009,strack_dicke_2011,baumann_dicke_2010,baumann_exploring_2011}. The key feature of these systems is the interplay between forces arising from photons scattered into the cavity by the atoms and the position-dependent atom-cavity coupling strengths. In the case of single-mode cavities, for example, the atomic spatial configuration undergoes a spontaneous symmetry breaking into one of two states dictated by the cavity boundary conditions~\cite{domokos_collective_2002,black_observation_2003}. The formation of more exotic phases of matter, such as supersolids and quantum spin glasses, has been predicted~\cite{gopalakrishnan_emergent_2009,strack_dicke_2011} and observed~\cite{baumann_dicke_2010,baumann_exploring_2011} by exploiting spatial interference between cavity and external pump fields or multiple cavity modes.

In this Letter, we show that atoms coupled to nanophotonic systems constitute a versatile platform for investigating similar behavior in the absence of externally imposed trapping potentials and boundary conditions set by cavity mirrors. This exciting frontier is motivated by recent experimental success in coupling cold atomic gases to nanophotonic waveguides~\cite{nayak_optical_2007,vetsch_optical_2010,goban_demonstration_2012}, and by predictions that certain atomic spatial configurations can yield remarkable optical response~\cite{kien_cooperative_2008,zoubi_hybrid_2010,chang_multiatomic_2011,chang_cavity_2012}, such as highly-reflecting atomic mirrors~\cite{chang_multiatomic_2011,chang_cavity_2012}. Whereas this previous work assumed the atoms were spatially fixed, here we show that the atoms can in fact self-organize into exotic spatial configurations due to the interplay of atomic internal states, positions, and photon-mediated forces~\cite{asboth_optomechanical_2008}.

\begin{figure}[p]
\begin{center}
\includegraphics[width=8cm]{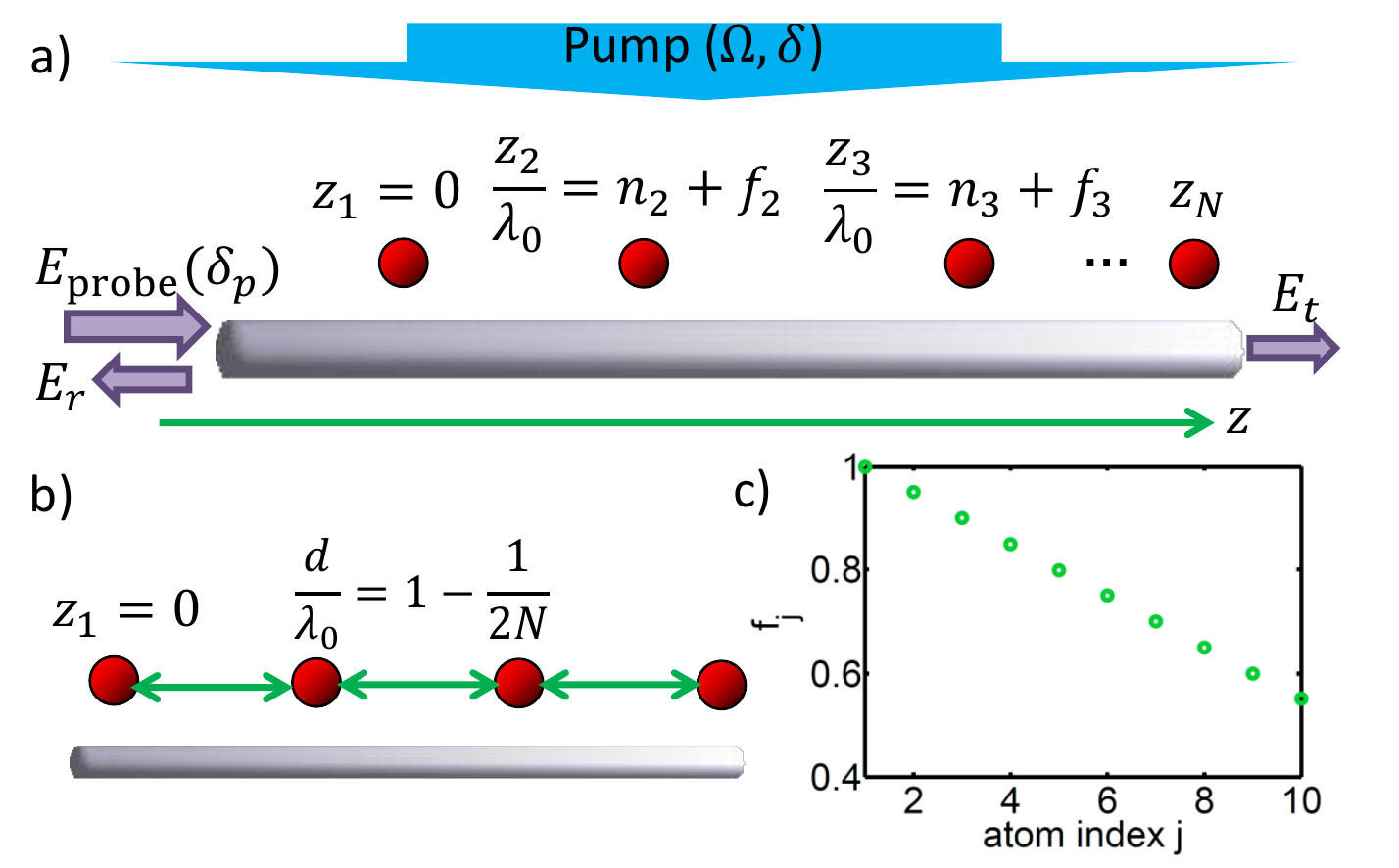}
\end{center}
\caption{a) Schematic illustration of $N$ atoms coupled to a single-mode waveguide, taken here to be an optical nanofiber, with physical positions along the waveguide $z_1<z_2<...<z_N$. The atoms are driven identically by a uniform pump field with Rabi frequency $\Omega$ and detuning $\delta$. The atomic positions $z_j/\lambda_0=n_j+f_j$, in units of the resonant wavelength $\lambda_0$, are parameterized by an integer $n_j$ and fraction $0<f_j\leq 1$. The system is fully characterized by $f_j$, as the integers $n_j$ have no physical consequence on the dynamics due to periodic, infinite-range interactions between atoms. A weak field of detuning $\delta_p$ incident through the waveguide separately probes the atomic configurations, \textit{e.g.}, via reflectance.  b) In the ``weak-scattering'' limit, the minimum energy state corresponds to an atomic lattice with lattice constant $d/\lambda_0=1-1/2N$ and fractional distances $f_j=1-(j-1)/2N$, as illustrated versus atom index $j$ in c) for the case of $N=10$ atoms.\label{fig:system}}
\end{figure}

Our system model generally describes an ensemble of $N$ atoms with ground and excited states $\ket{g},\ket{e}$, transition frequency $\omega_0$, and positions $z_j$~($1\leq j \leq N$), which experience dipole coupling with equal strength to a single-mode waveguide with left- and right-propagating modes~(see Fig.~\ref{fig:system}a). The atoms interact with each other via photons scattered into the waveguide modes, and effectively eliminating the associated fields produces a quantum spin model that describes atomic dipole-dipole interactions~\cite{chang_cavity_2012}. In particular, the reduced atomic density matrix obeys $\dot{\rho}=-i[H_{\footnotesize\textrm{dd}},\rho]+\mathcal{L}_{\footnotesize\textrm{dd}}[\rho]$, where
\bea H_{\footnotesize\textrm{dd}} & = & \frac{\GammaD}{2}\sum_{j,j'}\sin\,k_{0}|z_j-z_{j'}|\sigma_{eg}^{j}\sigma_{ge}^{j'},\label{eq:Hdd} \\ \mathcal{L}_{\footnotesize\textrm{dd}}[\rho] & = & \sum_{l=\pm}2\hat{\mathcal{O}}_l \rho \hat{\mathcal{O}}_l^{\dagger} -\hat{\mathcal{O}}_l^{\dagger}\hat{\mathcal{O}}_l\rho -\rho\hat{\mathcal{O}}_l^{\dagger}\hat{\mathcal{O}}_l. \eea
Here, the jump operators are defined by $\hat{\mathcal{O}}_\pm=\sqrt{\GammaD/4}\sum_{j}\sigma_{ge}^{j}e^{\mp ik_{0}z_j}$, where $\GammaD$ denotes the spontaneous emission rate of a single, independent atom into the guided modes, $k_0$ is the wavevector at the atomic resonance frequency, and $\sigma_{ge}^j=\ket{g^j}\bra{e^j}$ is the lowering operator for atom $j$. Physically, $H_{\footnotesize\textrm{dd}}$ describes coherent dipole-dipole coupling between atoms $j,j'$, while $\mathcal{L}_{\footnotesize\textrm{dd}}[\rho]$ captures cooperative emission~(such as super- and subradiance~\cite{gross_superradiance:_1982}) mediated by the two guided directions~($l=\pm$). An important feature of our 1D system is that the dipole interactions involve all pairs of atoms, and are approximately infinite in range and oscillatory in strength. This reflects the capability of the waveguide to channel a photon emitted by one atom without diffraction to a distant second atom, with the interaction depending only on the relative phase between the atomic coherences. Corrections from causality and free-space propagation phase~(the derivation assumes that all relevant wavevectors are close to $k_0$) are negligible for realistic system sizes and atom number $N$~\cite{chang_cavity_2012}. However, we must account for a non-negligible spontaneous emission rate $\Gamma'$ into free space, which we include via additional Liouvillian terms with independent jump operators $\hat{\mathcal{O}}_{j}'=\sqrt{\Gamma'/2}\sigma_{ge}^j$~($1 \leq j \leq N$).

We emphasize that photon-mediated forces can be extremely large in nanophotonic systems, which allows their effects to be prominent. Specifically, the strength $\GammaD$ of the pairwise interaction in Eq.~(\ref{eq:Hdd}) can be a significant fraction of the vacuum emission rate $\Gamma_0$, which corresponds to the Doppler temperature in atomic motion.  As shown below, the long-range nature of interactions can produce forces further enhanced by a factor $\sim N$ over a single pair. Current systems with nanofibers attain coupling strengths approaching $\GammaD\sim 0.1\Gamma_0$~\cite{nayak_optical_2007,vetsch_optical_2010,goban_demonstration_2012}, with further increases expected from improved waveguide design and trapping techniques.

Now we consider the case where the atoms are driven identically by a pump field with Rabi frequency $\Omega$ and detuning $\delta=\omega_{\footnotesize\textrm{pump}}-\omega_0$, as illustrated in Fig.~\ref{fig:system}a. We focus on the semi-classical limit, in which the motion of the atoms is treated classically and atomic saturation is ignored. The latter assumption allows for different atoms to be de-correlated, \textit{e.g.}, $\avg{\sigma^j\sigma^{j'}}\approx\avg{\sigma^j}\avg{\sigma^{j'}}$. The evolution equations of the operator mean values satisfy
\bea \dot{z}_j & = & p_j/m,\label{eq:zdot} \\ \dot{\sigma}_{ge}^j & = &  (i\delta-\Gamma/2)\sigma_{ge}^j+i\Omega-\frac{\GammaD}{2}\sum_{j'\neq j}\sigma_{ge}^{j'}e^{ik_0|z_j-z_{j'}|},\label{eq:sigmadot} \\ \dot{p}_j & = & -(\hbar k_0)\GammaD \textrm{Re}\left[\sum_{j'}\sigma_{ge}^j \sigma_{eg}^{j'}e^{-ik_0|z_j-z_{j'}|}\textrm{sign}(z_j-z_{j'})\right].\label{eq:pdotcl} \eea
For notational simplicity, we will avoid the explicit use of brackets $\avg{}$ to denote the mean values. Here $\Gamma=\GammaD+\Gamma'$ is the total spontaneous emission rate of an independent atom, $m$ is the atomic mass, and $p_j$ is the momentum of atom $j$. An equivalent formulation of the optical forces based upon transfer matrices was derived in Ref.~\cite{asboth_optomechanical_2008}, which investigated the formation of 1D optical lattices given asymmetric trapping fields.

Eqs.~(\ref{eq:zdot})-(\ref{eq:pdotcl}) determine the full atomic dynamics and the possibility of self-organized configurations. We now briefly summarize our approach to finding these solutions. A configuration is self-organizing if it is a steady-state solution to Eqs.~(\ref{eq:zdot})-(\ref{eq:pdotcl}) and if this solution is stable with respect to small perturbations. In the ``weak-scattering'' limit of large pump detuning or low optical depth, we show that there is a unique solution that satisfies an energy minimization condition. Generally, however, it is not possible to formulate the problem in terms of energy minimization. Thus, we employ an adiabatic procedure, in which the weak-scattering solution at large detuning is continuously transformed by bringing the detuning toward resonance in small steps and finding the new stable configuration each time. Numerically, we accomplish this by introducing a small, external momentum damping rate for each atom and integrating Eqs.~(\ref{eq:zdot})-(\ref{eq:pdotcl}) for each new detuning until the system reaches steady state. Such an external damping term is in fact needed to achieve true steady-state behavior, to compensate for internal anti-damping forces arising from the delayed response of the atomic coherences to the motion. In practice, however, we find that anti-damping can be made negligible over experimentally relevant time scales.

To describe the self-organized configurations, we assume without loss of generality that $z_j\leq z_{j'}$ for $j<j'$ and define $z_1=0$. While the forces in Eq.~(\ref{eq:pdotcl}) only influence the relative coordinates and the center-of-mass motion of the atoms in principle remains free, in practice this motion will have nearly zero velocity due to the initial cooling required to load atoms along the waveguide~\cite{vetsch_optical_2010,goban_demonstration_2012}. It is convenient to write the positions in the form $z_j/\lambda_0=n_j+f_j$, where $n_j$ is an integer and $f_j$ is a fractional distance~($0<f_j\leq 1$), as shown in Fig.~\ref{fig:system}a. Due to the periodic nature of Eqs.~(\ref{eq:sigmadot}) and~(\ref{eq:pdotcl}), the integers $n_j$ have no observable consequence in the evolution, and the system is fully characterized by $f_j$.

The ``weak-scattering'' regime is defined as that where the fields re-scattered by the atoms into the waveguide~(proportional to the last term on the right-hand side in Eq.~(\ref{eq:sigmadot})) are negligible compared to the external driving field. A sufficient condition is $N\GammaD\ll\sqrt{\delta^2+(\Gamma/2)^2}$, in which case the atomic coherences are equal and given by $\sigma_{ge}^j\approx\sigma_{ge}^{(0)}\equiv i\Omega/(\Gamma/2-i\delta)$. The atomic motion is then governed by a purely mechanical potential, $H_{\footnotesize\textrm{dd}}\approx(\GammaD s_0/2)\sum_{j,j'}\sin k_{0}|z_j-z_{j'}|$, where $s_0\equiv |\sigma_{ge}^{(0)}|^2$. This potential is minimized when the atoms form a lattice with lattice constant $d_{\footnotesize\textrm{ws}}=\lambda_0(1-\frac{1}{2N})$~(see Fig.~\ref{fig:system}b), with corresponding energy $H_{\footnotesize\textrm{dd,min}}=-N^2\frac{\GammaD s_0}{\pi}$. The $N^2$ scaling reflects the strong collective forces arising from the long-range interactions within the system. The fractional distances take the form $f_j=1-(j-1)/2N$, as illustrated in Fig.~\ref{fig:system}c for the case of $N=10$ atoms.

An atomic lattice spaced by the resonant wavelength $\lambda_0=2\pi/k_0$ forms a perfect mirror on resonance as $N\rightarrow\infty$, even though a single atom is mostly absorptive~\cite{chang_cavity_2012}. This phenomenon arises due to the coupling of light to a superradiant spin wave $\hat{S}=\sum_{j}\sigma_{ge}^j$ of the atoms. As described below, for large $N$ the close proximity of the weak-scattering configuration~(with lattice constant $d_{\footnotesize\textrm{ws}}=\lambda_0(1-\frac{1}{2N})$) to the superradiant one gives rise to intriguing crossover behavior as the pump frequency is tuned closer to resonance~($\delta\rightarrow 0$) and the atomic optical depth increases.

\begin{figure}[p]
\begin{center}
\includegraphics[width=14cm]{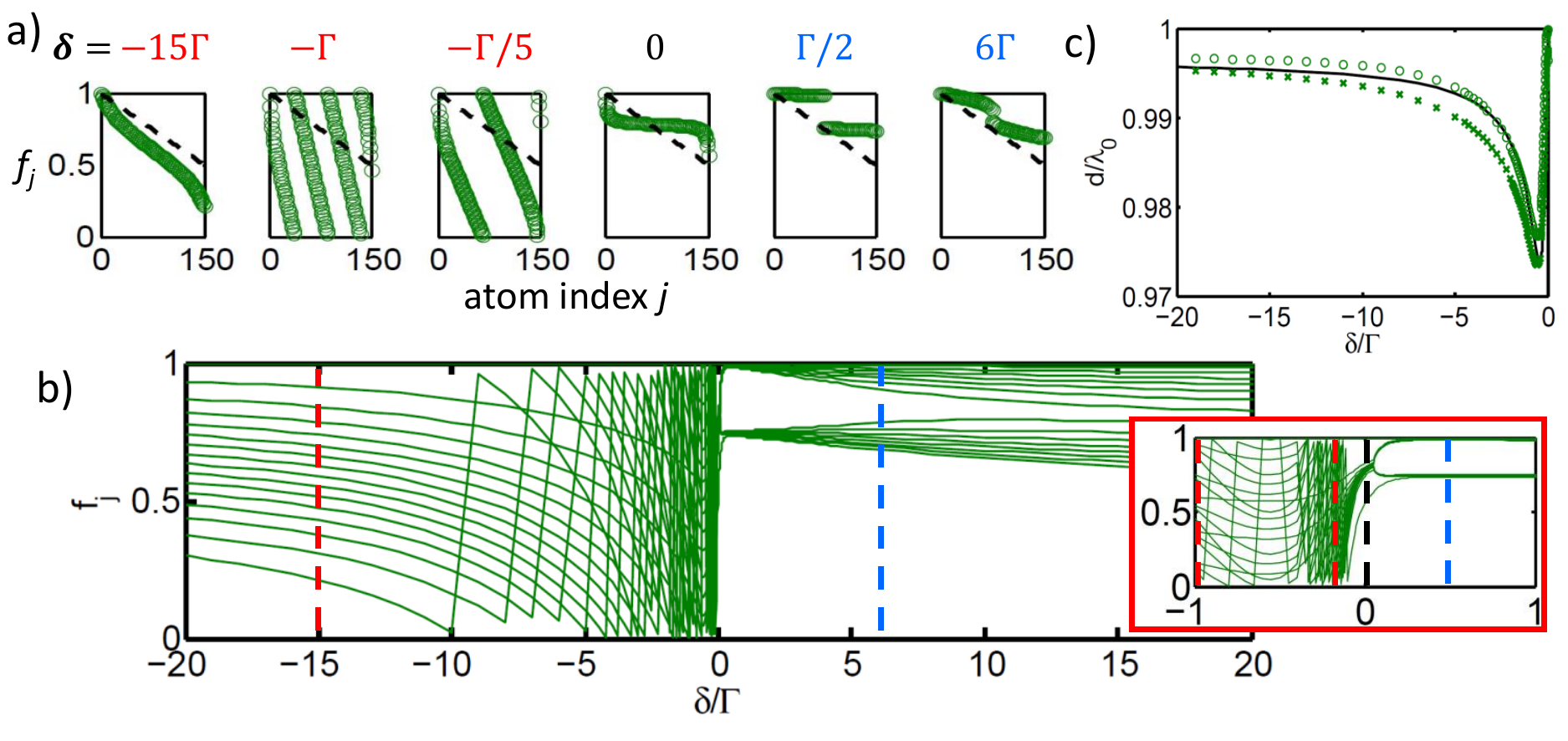}
\end{center}
\caption{a) Fractional positions $f_j$ versus atom index for selected pump detunings $\delta/\Gamma=-15,-1,-0.2,0,0.5,6$. All numerical simulations are for $N=150$ atoms and $\GammaD=\Gamma/4$. The dotted line represents the solution in the weak-scattering limit. b) Positions $f_j$ versus pump detuning. The vertical dashed lines denote the detunings for which the positions are plotted in Fig.~\ref{fig:spatial}a. For clarity, only 16 representative atomic positions are shown here. Inset: zoom of the same plot near resonance. c) Characteristic lattice constant $d/\lambda_0$ versus detuning for $\delta<0$. The circles and crosses denote the lattice constant as determined by the central two atoms and the average over all atoms, respectively. The solid curve depicts $d_{\footnotesize\textrm{eff}}=\lambda_{\footnotesize\textrm{eff}}(1-1/2N)$, where $\lambda_{\footnotesize\textrm{eff}}$ accounts for the effective refractive index provided by the atomic medium.\label{fig:spatial}}
\end{figure}

Starting from the weak-scattering solution at large detuning, we obtain adiabatically transformed solutions by changing $\delta$ in small steps and integrating Eqs.~(\ref{eq:zdot})-(\ref{eq:pdotcl}) at each step until the system converges to a stationary state. A small external damping $\dot{p}_j=-\gamma_e p_j$ is added to the equations to facilitate convergence. Fig.~\ref{fig:spatial}a depicts some self-organization solutions $f_j$ for a representative set of detunings, in the case of $N=150$ atoms and $\GammaD=\Gamma/4$~(also see Fig.~\ref{fig:spatial}b depicting continuous variation of $\delta$). Compared to the weak-scattering regime, these solutions exhibit a variety of interesting phenomena including lattice compression, expansion, and phase slips or fragmentation. While we have chosen a particular parameter set for illustration here, we emphasize that our conclusions are quite general.

We first consider the approach toward resonance~($\delta=0$) starting from large negative detuning. Our calculations reveal a gradual compression of the lattice constant, $d<d_{\footnotesize\textrm{ws}}$, as $\delta/\Gamma\rightarrow-1/2$, followed by a rapid expansion between $-1/2\lesssim\delta/\Gamma\lesssim 0$~(Fig.~\ref{fig:spatial}c) and a crossover into the superradiant regime $d\approx\lambda_0$~(nearly equal $f_j$). A simple model for moderate scattering strength can be derived by noting that the atoms primarily act dispersively for large detunings. The phase shift imparted in transmission by a single atom can readily be calculated~\cite{chang_single-photon_2007,chang_cavity_2012}, $\theta_t=-\arctan\left(\frac{2\GammaD\delta}{\Gamma^2-\Gamma\GammaD+4\delta^2}\right)$, which leads to an effective frequency-dependent wavelength $\lambda_{\footnotesize\textrm{eff}}\approx\lambda_0(1-\theta_t/2\pi)$. For negative detuning, $\lambda_{\footnotesize\textrm{eff}}<\lambda_0$~(\textit{i.e.}, the medium has an effective index $n_{\footnotesize\textrm{eff}}>1$), which accounts for the smaller lattice constant $d<d_{\footnotesize\textrm{ws}}$. Furthermore, the dispersive shift is maximized when $\delta/\Gamma\approx-1/2$. In Fig.~\ref{fig:spatial}c, we plot the predicted lattice constant $d_{\footnotesize\textrm{eff}}=\lambda_{\footnotesize\textrm{eff}}(1-1/2N)$ in the regime $\delta<0$, as compared to the numerically obtained lattice constants. Our simple model reproduces well the actual behavior, even for large optical depths when one expects the system to be better described by band structure than an effective index~\cite{deutsch_photonic_1995}.

\begin{figure}[p]
\begin{center}
\includegraphics[width=9cm]{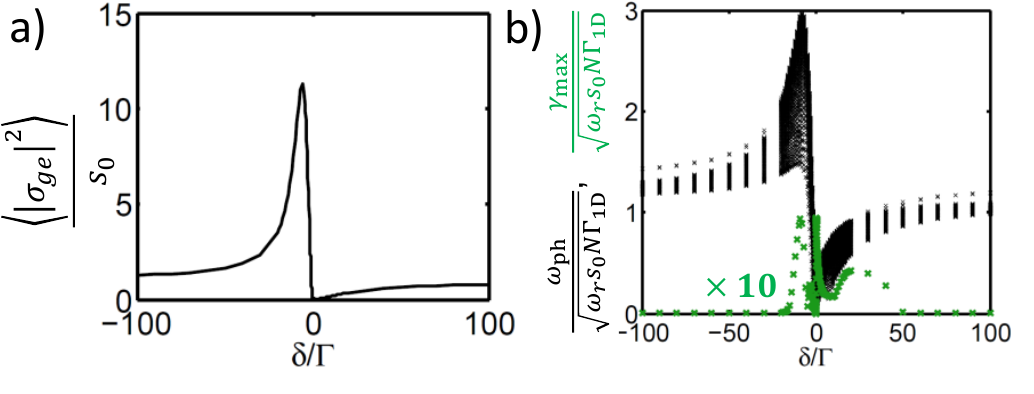}
\end{center}
\caption{a) Mean excited state population $\avg{|\sigma_{ge}|^2}$ of the atoms versus dimensionless pump detuning $\delta/\Gamma$, for the system considered in Fig.~\ref{fig:spatial}. The population is normalized by the result $s_0=\frac{\Omega^2}{\delta^2+(\Gamma/2)^2}$ for an independently driven atom. b) Black crosses: phonon frequencies $\omega_{\footnotesize\textrm{ph}}$ of the relative motional modes in the self-organization configurations, as a function of $\delta/\Gamma$. Green crosses: maximum anti-damping rate $\gamma_{\footnotesize\textrm{max}}$ of the motional modes, scaled up by a factor of $10$ for better contrast. Both $\omega_{\footnotesize\textrm{ph}}$ and $\gamma_{\footnotesize\textrm{max}}$ are plotted in units of $\sqrt{\omega_r s_0 N \GammaD}$. We have used $\omega_r=10^{-3}\Gamma$ for these calculations.\label{fig:phonons}}
\end{figure}

Approaching resonance from large positive detuning, a similar argument would predict lattice expansion, with the system attaining the superradiant lattice constant of $d=\lambda_0$ at a critical detuning of $\delta_c\sim N\GammaD/2\pi$. Such an argument is inconsistent, however, as the atoms do not behave as independent refractive index elements in the superradiant regime. Instead, we find that the system crosses over into a new regime around $\delta\sim\delta_c$, where the single lattice fragments into two smaller lattices with a ``phase slip'' between them~(\textit{e.g.}, see Fig.~\ref{fig:spatial}a for detuning $\delta=\Gamma/2$). In particular, the left and right halves of the atomic system have approximately constant  $f_j$~(indicating a lattice constant in those segments of $\sim\lambda_0$), while a phase slip of $\Delta f \approx-1/4$ occurs between those two segments~(corresponding to a $3\lambda_0/4$ separation between the two halves). Physically, each segment behaves like a collective ``super-atom'' with large dipole coupling to the guided modes. These two super-atoms are bound together by optical forces in the same way that two single atoms would self-organize. Specifically, the spacing between the two super-atoms is expected to minimize a reduced two-particle potential $H_{\footnotesize\textrm{dd}}\propto\sin k_{0}|z_R-z_L|$, which occurs for $z_R-z_L=3\lambda_0/4$ and reproduces the observed phase slip. The association of this behavior with superradiance can also be seen in Fig.~\ref{fig:phonons}a. Here, we have plotted the mean atomic excited state population normalized by the independent atom result, $\avg{|\sigma_{ge}|^2}/s_0$, as a function of detuning. For positive detunings, the excited state population is strongly suppressed due to the large collective decay rate into the guided modes.

Near resonance, we find that the phase slip rapidly vanishes and the two segments merge to again form a single lattice. Thus far, we are not able to produce a simple effective model of this phenomenon, as the strong, long-range interactions in this regime cause simple ``mean-field'' descriptions involving just a few parameters to apparently break down. However, we believe that the nature of this many-atom, strongly interacting system will be a rich area for future investigations.

For small displacements around the self-organized solutions, the atomic motion is well-described by a set of normal modes~(\textit{i.e.}, phonons). A good description of these modes can be obtained by first noting that the ratio of the atomic recoil frequency $\omega_r=(\hbar k_0^2)/2m$ to the spontaneous emission rate $\Gamma$ is typically small. This implies that the atomic coherences $\sigma_{ge}$ follow the motion nearly instantaneously. Writing Eq.~(\ref{eq:sigmadot}) in the form $\dot{\sigma}_{ge}^j=M_{jk}(\vec{z}(t))\sigma_{ge}^k+i\Omega$, the lowest-order solution is given by
\be \sigma_{ge,\footnotesize\textrm{inst}}=-iM^{-1}(\vec{z}(t))\vec{\Omega}.\label{eq:sigmainst} \ee
Substituting this expression back into the force equation~(\ref{eq:pdotcl}) and linearizing around the equilibrium positions $\vec{z}_{\footnotesize\textrm{eq}}$ yields a set of restoring equations $\dot{p}_j=-K_{jk}(z_k-z_{\footnotesize\textrm{eq},k})$.

In the weak-scattering limit, $K_{jk}$ is a circulant matrix, which enables the phonon spectrum to be solved exactly. Specifically, we find $N$ phonon modes with wave numbers $2\pi j/N$~($0\leq j \leq N-1$) and frequencies
\be \omega_{\footnotesize\textrm{ph},j}=\left[2\omega_r s_0 \GammaD \left(\cot \frac{\pi}{2N}-\frac{\sin (\pi/N)}{\cos (2\pi j/N)-\cos (\pi/N)}\right)\right]^{1/2}. \ee
The trivial $j=0$ mode denotes the free~(zero-frequency) center-of-mass motion. The other modes scale with large $N$ approximately as $\omega_{\footnotesize\textrm{ph},j}\sim \sqrt{\omega_r s_0 N \GammaD}$. In Fig.~\ref{fig:phonons}b, we plot the normalized phonon frequencies $\omega_{\footnotesize\textrm{ph},j}/\sqrt{\omega_r s_0 N \GammaD}$ as a function of detuning. The phonon spectrum exhibits a strong softening in the superradiant regime~(Fig.~\ref{fig:phonons}a). It should be noted that our approach to find self-organized solutions based upon adiabatic transformation does not preclude the possibility of other solutions. In fact, preliminary calculations suggest that the phonon softening enables the system to develop a multitude of other weakly stable solutions. This phenomenon will be studied more extensively in future work.

Small delays in the atomic internal response compared to the motion can lead to damping or anti-damping forces. We can characterize this effect perturbatively by substituting Eq.~(\ref{eq:sigmainst}) into~(\ref{eq:sigmadot}) and iteratively finding higher-order corrections. In particular, we can write $\sigma_{ge}(t)\approx \sigma_{ge,\footnotesize\textrm{inst}}+\sigma_{ge,\footnotesize\textrm{d}}$, where the first-order correction satisfies
\be \sigma_{ge,\footnotesize\textrm{d}}=M^{-1}(\vec{z}_{eq})\frac{\vec{p}\cdot\nabla}{m}\sigma_{ge,\footnotesize\textrm{inst}}(\vec{z}_{eq}). \ee
Like before, this expression can be substituted into Eq.~(\ref{eq:pdotcl}) to yield $\dot{p}_j=-K_{jk}(x_k-x_{eq,k})-L_{jk}p_k$, where the matrix $L_{jk}$ characterizes momentum damping or anti-damping. This term produces an imaginary component in the normal mode frequencies, $\omega_{\footnotesize\textrm{ph},j}\rightarrow \omega_{\footnotesize\textrm{ph},j}+i\gamma_{\footnotesize\textrm{ph},j}$, where $\gamma_{\footnotesize\textrm{ph},j}>0$~($\gamma_{\footnotesize\textrm{ph},j}<0$) indicates anti-damping~(damping). For sufficiently large atom number $N$, we generally find that some of the modes exhibit anti-damping. In the weak-scattering limit, the largest anti-damping rate behaves approximately as $\gamma_{\footnotesize\textrm{max}}\sim \frac{N^2\GammaD^2 s_0 \omega_r}{\delta^2}$. While anti-damping in principle implies that the system is only meta-stable absent an external cooling mechanism, in practice, the anti-damping rates can be made negligible compared to typical inverse trapping lifetimes for cold atoms. In the weak-scattering limit~($N\GammaD\ll|\delta|$), for example, $\gamma_{\footnotesize\textrm{max}}\ll s_0\omega_r$ is a small fraction of the recoil frequency. In Fig.~\ref{fig:phonons}b, we plot the maximum anti-damping rate $\gamma_{\footnotesize\textrm{max}}$ versus detuning.

Self-organization can be observed through its dramatic influence on light propagation through the waveguide. Here, we consider the linear reflection and transmission of an incident guided probe beam, as illustrated in Fig.~\ref{fig:system}a. For concreteness, we do not account for the atomic forces or motion imparted by the probe. For example, once the atoms evolve to the steady-state, self-organized positions as in Figs.~\ref{fig:spatial}a,b, the resulting configuration would be probed in a destructive fashion by a probe pulse $E_{\footnotesize\textrm{probe}}(t)$ of duration $\tau \ll \omega_{\footnotesize\textrm{ph}}$. As discussed in Refs.~\cite{chang_cavity_2012,chang_single-photon_2007}, the reflection and transmission coefficients of a single atom are given by $r(\delta_p)=-\frac{\GammaD}{\Gamma-2i\delta_p}$ and $t(\delta_p)=1+r(\delta_p)$, where $\delta_p$ is the probe detuning. Given the coefficients of a single atom, one can efficiently calculate the reflection and transmission amplitudes of an array~(including free-space propagation between atoms) via the transfer matrix technique~\cite{asboth_optomechanical_2008,deutsch_photonic_1995}.

It is helpful to first consider light propagation through an infinite, perfect lattice with lattice constant $d$, in which case the optical modes are well-described by Bloch wavevectors~\cite{joannopoulos_photonic_2008}. Following the techniques of Ref.~\cite{deutsch_photonic_1995}, we find that the optical Bloch wavevectors $q$ obey the dispersion relation $\cos qd \approx \cos k_{0}d-\zeta\sin k_{0}d$, where $\zeta=\frac{\GammaD}{\Gamma'}\frac{i-2\delta_p/\Gamma'}{1+(2\delta_p/\Gamma')^2}$. Furthermore, in analogy with purely dispersive media, we are motivated to temporarily ignore the atomic absorption (imaginary part of $\zeta$) and look for ``band gap'' regions~(where $q$ is purely imaginary), in which propagation is forbidden due to strong interference in the multiple reflections. Defining a parameter $\epsilon=2\pi(1-d/\lambda_0)$ that characterizes the offset from the superradiant lattice constant $d=\lambda_0$, we find that an optical band gap exists for probe detunings $-\frac{\GammaD}{\epsilon} \lesssim \delta_{\footnotesize\textrm{gap}} \lesssim -\frac{\epsilon\Gamma'^2}{4\GammaD}$~(assuming that $\GammaD/\Gamma'\gg\epsilon$). In dispersive media, a band gap gives rise to perfect reflection from a long lattice due to the absence of propagating modes within the system.

\begin{figure}[p]
\begin{center}
\includegraphics[width=9cm]{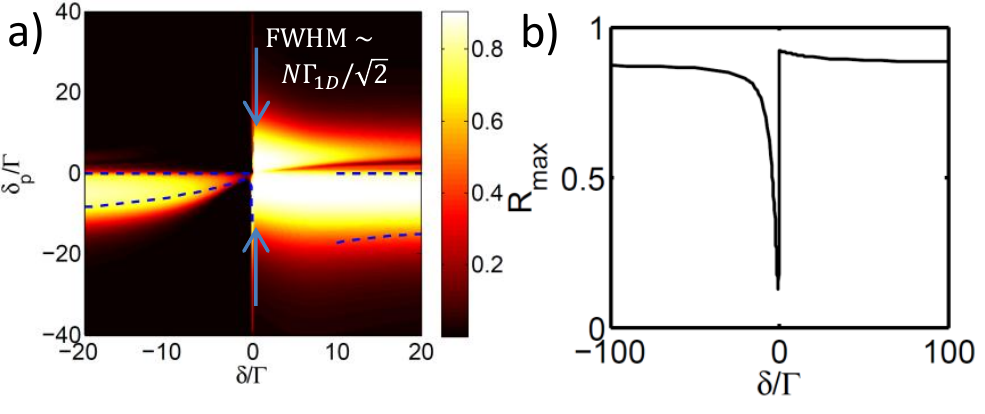}
\end{center}
\caption{a) Reflectance from self-organized atomic configuration, versus pump~($\delta$) and probe~($\delta_p$) detunings. The blue dashed curves indicate the edges of the band gap region $\delta_{\footnotesize\textrm{gap}}$, which is well-defined when a single lattice constant exists~(\textit{i.e.}, outside of the phase slip configuration). The lattice constant used here is the average over all atoms. For the phase slip configuration, the reflectance spectrum is nearly symmetric with respect to $\delta_p$, with a FWHM given by $\sim N\GammaD/\sqrt{2}$. b) Maximum reflectance (optimized with respect to $\delta_p$) versus pump detuning. All simulations are for $N=150$ atoms and $\GammaD/\Gamma=0.25$.\label{fig:spectra}}
\end{figure}

Self-organization in our system into a lattice with $d<\lambda_0$~($\epsilon>0$) thus should manifest itself in an asymmetric peak in the reflectance spectrum, as the band gap occurs for negative probe detunings $\delta_{p}<0$. The width of this reflection peak should scale approximately like $\epsilon^{-1}$. In Fig.~\ref{fig:spectra}a, we have plotted the reflection spectra for the different self-organization configurations~(as determined by the pump detuning $\delta$). Here, we have evaluated the spectra based on the numerical solutions of Fig.~\ref{fig:spatial}b and including fully atomic absorption. We have overlaid the edges of the band gap $-\frac{\GammaD}{\epsilon} \lesssim \delta_{\footnotesize\textrm{gap}} \lesssim -\frac{\epsilon\Gamma'^2}{4\GammaD}$~(dashed blue curves) in the regimes where the system is well-described by a single lattice constant~(\textit{i.e.}, outside of the phase slip configuration). A clear correlation between enhanced reflectance and the predicted band gap is observed, with the absence of perfect reflectance attributable to the finite atom number, atomic absorption, and small variations of $\epsilon$ along the atomic chain.

In the phase slip configuration that occurs for positive pump detuning, the reflection spectra change dramatically. In analogy with Ref.~\cite{chang_cavity_2012}, this system can be modeled as two high-reflectivity atomic mirrors forming a high-finesse cavity mode, which enables its optical properties to be easily determined. Here, the spectrum becomes symmetric with respect to $\delta_p$, with the peak reflectance and full-width given respectively by $R\approx 1-4\Gamma'/(N\GammaD)$ and $\delta_{\footnotesize\textrm{FWHM}}\approx N\GammaD/\sqrt{2}$. The transition from asymmetric to symmetric spectra as the atoms cross over to the phase slip configuration is evident in Fig.~\ref{fig:spectra}a. In Fig.~\ref{fig:spectra}b, we have plotted the peak reflectance~(maximized over probe detuning $\delta_p$) for each spatial configuration determined by the pump detuning $\delta$.

In summary, we have shown that cold atoms coupled to nanophotonic waveguides can exhibit rich self-organization behavior due to the interplay between atomic motion, internal states, and optical response, when the dynamics are treated in the semi-classical regime. However, we believe that a variety of novel quantum phenomena should arise as well. For example, it would be interesting to investigate how the optical fields become entangled with the atomic positions~\cite{cormick_structural_2012}, or account for correlations that would reveal the emergence of quantum phases of matter~\cite{strack_dicke_2011}. Furthermore, it should be possible to tailor more exotic spin-force Hamiltonians by exploiting multi-level atomic structure~(see, \textit{e.g.}, Ref.~\cite{chang_cavity_2012} for a $\Lambda$ configuration) and by using photonic crystals~\cite{joannopoulos_photonic_2008} to engineer the dimensionality and geometry of the atom-nanophotonics interfaces. Beyond the exploration of many-body phenomena, we envision that these systems can also find applications in quantum information processing or few-photon nonlinear optical devices~\cite{greenberg_high-order_2012}. For example, it may be possible to use phonons in self-organized configurations to manipulate and transfer quantum information over long distances, much like in ion trap systems~\cite{leibfried_quantum_2003}, or to exploit self-organization arising from an initial pump pulse to exert a nonlinear response on a subsequent probe pulse.

The authors thank O. Painter and H. Ritsch for helpful discussions. DEC acknowledges support from Fundaci\'{o} Privada Cellex Barcelona. HJK acknowledges funding from the IQIM, an NSF Physics Frontier Center with support of the Moore Foundation, by the AFOSR QuMPASS MURI, by the DoD NSSEFF program, and by NSF PHY-1205729. JIC acknowledges funding from the EU Project AQUTE.


\bibliographystyle{apsrev_nourl}
\bibliography{self-organization_bib}

\end{document}